\PassOptionsToPackage{pdfpagelabels=false}{hyperref} 
\documentclass[fleqn,usenatbib]{mnras}

\usepackage{newtxtext,newtxmath}

\usepackage[T1]{fontenc}

\makeatletter
\newlength{\abovecaptionskip}%
\makeatother

\usepackage{amsmath,amstext}
\usepackage{graphicx}	
\usepackage[figure,figure*]{hypcap}
\usepackage{enumerate}
\usepackage{xspace} 
\usepackage{multirow}
\usepackage[flushleft]{threeparttable}
\usepackage{booktabs}
\usepackage{adjustbox}
\usepackage{marginnote}

\usepackage{soul}
\setstcolor{red}

\makeatletter
\newcommand{\HI}{{\text{H\,\MakeUppercase{\small\romannumeral 1}}}\xspace}
\newcommand{\HII}{{\text{H\,\MakeUppercase{\small\romannumeral 2}}}\xspace}

\newcommand{\HeII}{{\text{He\,\MakeUppercase{\small\romannumeral 2}}}\xspace}

\newcommand{\Lya}{\ifmmode{\mathrm{Ly}\alpha}\else Ly$\alpha$\xspace\fi}
\newcommand{\Htwo}{\ifmmode{\mathrm{H}_2}\else H$_2$\xspace\fi}
\newcommand{\MBII}{{\text{MB\MakeUppercase{\small\expandafter{\romannumeral 2}}}}\xspace}

\newcommand{\Rmnum}[1]{\expandafter\@slowromancap\romannumeral #1@}

\newcommand{\Msun}[1]{\ensuremath{{\rm M}_{\odot}} #1 }

\newcommand{\abb}[2]{{#1}{\small \uppercase\expandafter{\romannumeral #2}}}
\newcommand{\abbm}[2]{\ensuremath{{\rm #1}{\small\rm\scriptstyle \uppercase\expandafter{\romannumeral #2}}}}

\title[Faint Black Holes]{Mocking Faint Black Holes during Reionization}

\author[M.B.~Eide et al.]{Marius B. Eide$^{1}$\thanks{E-mail: \href{mailto:eide@mpa-garching.mpg.de}{eide@mpa-garching.mpg.de}}
Benedetta Ciardi$^{1}$, 
Yu Feng$^{2}$
and Tiziana Di Matteo$^{3,4}$
\\
$^{1}$Max-Planck-Institut f\"ur Astrophysik, Karl-Schwarzschild-Stra\ss e 1, 85741 Garching, Germany\\
$^{2}$Berkeley Center for Cosmological Physics Campbell Hall 341, University of California, Berkeley CA 94720, United States\\
$^{3}$McWilliams Center for Cosmology, Physics Department, Carnegie Mellon University, Pittsburgh, PA 15213, USA\\
$^{4}$School of Physics, The University of Melbourne,
VIC 3010, Australia
\\
}

\date{Accepted 2020 October 15. Received 2020 September 23; in original form 2020 July 11}

\pubyear{2020}

\begin{document}
\label{firstpage}
\pagerange{\pageref{firstpage}--\pageref{lastpage}}
\maketitle

\begin{abstract}
To investigate the potential abundance and impact of nuclear black holes (BHs) during reionization, we generate a neural network that estimates their masses and accretion rates by training it on 23 properties of galaxies harbouring them at $z=6$ in the cosmological hydrodynamical simulation Massive-Black II. We then populate all galaxies in the simulation from $z=18$ to $z=5$ with BHs from this network. As the network allows to robustly extrapolate to BH masses below those of the BH seeds, we predict a population of faint BHs with a turnover-free luminosity function, while retaining the bright (and observed) BHs, and together they predict a Universe in which intergalactic hydrogen is $15\%$ ionized at $z=6$ for a clumping factor of 5. Faint BHs may play a stronger role in H reionization without violating any observational constraints. This is expected to have an impact also on pre-heating and -ionization, which is relevant to observations of the 21~cm line from neutral H. We also find that BHs grow more efficiently at higher $z$, but mainly follow a redshift-independent galaxy-BH relation. We provide a power law parametrisation of the hydrogen ionizing emissivity of BHs.
\end{abstract}


\section{Introduction}
\label{sec:intro}

Black holes (BHs) have been prime candidates for the ionization of the 
Universe  (e.g.~\citealt{Rees1969} and \citealt{Arons1969}) ever since 
the early days of the unavailing search for the intergalactic medium 
(IGM, e.g.~\citealt{Field1959}).
With the detection of 22 faint BHs at $z>4$, \cite{Giallongo2015}
revived the question of their role in this process.
With the optical depth of the intergalactic free electrons being as low as $\tau = 0.054 \pm 0.007$ \citep{PlanckCollaborationVI2018}, hydrogen reionization is expected to end late enough for stars to be its main driver \citep[e.g.~][]{Robertson2015,Bouwens2015ems}. Nevertheless, an observational picture where stars alone are responsible for reionization might need to rely on an escape fraction, $f_{\rm esc}$, of ionizing photons from galaxies during the epoch of reionization (EoR) higher than what observed at lower redshifts (see e.g.~\citealt{Naidu2018}, but note also the high individual $f_{\rm esc}$ found by e.g. \citealt{Vanzella2016} or \citealt{Fletcher2019}), as well as on a population of unobserved faint galaxies.
There is thus room for BHs even in the picture of a stellar-dominated EoR. The question remains how large their contribution is.

In the most extreme case, \cite{Madau2015} found a BH-only reionization scenario under the \cite{Giallongo2015} constraints to positively match the evolution of the volume filling factor of ionized hydrogen, $Q_\HII$. This scenario however fails at reproducing several other observations. BHs alone would yield IGM temperatures and heating that are too high \citep[see e.g.~the comparison of BH-only models to the compilation of IGM temperatures of~][]{Garaldi2019}, which is followed by too early adiabatic cooling. Furthermore, in the BHs dominated model of \cite{Madau2015}, \HeII reionization would be completed prematurely at $z \sim 4.2$, shortly after \HI reionization, which is at odds with the observed extended \HeII reionization process \citep{Worseck2016,Worseck2019}.
The observational constraints on the ionizing ouptut from high-$z$
\citep[e.g.~][]{Onoue2017,Parsa2018,Matsuoka2018,Kulkarni2019},
as well as theoretical inferences \citep[e.g.~][]{Finkelstein2019}, indicate that BHs supply a significant contribution to the ionizing budget, albeit subdominant to that of stars. As discussed by \cite{DAloisio2017}, BHs can provide an elegant explanation to a flat redshift evolution of the ionizing emissivity, justify the low optical depths in the \HeII Lyman $\alpha$ forest, and importantly, explain the origin for the large variations in the opacity of \HI Lyman $\alpha$ forest along different sightlines \citep[as investigated by e.g.~][]{Chardin2015}.

The interplay between BHs and their host galaxies shapes them both \citep[e.g.~][]{DiMatteo2005}.
Observations have revealed that massive BHs exist already by $z=7.5$ \citep{Banados2018,Fan2019}, and  simulations do not rule this out as unfeasible \citep[e.g.~][]{Feng2015,  DiMatteo2017}. The growth of BHs can be captured well by simulations \citep[e.g.~][]{Sijacki2015, Degraf2012, Weinberger2017, Huang2018}, however, the question of their formation remains still open \citep[e.g.;][for recent reviews]{Regan2009,Volonteri2012,Inayoshi2019}. A common numerical approach in large cosmological volume simulations \citep{DiMatteo2012, Khandai2015, Sijacki2015, Crain2015, Weinberger2017}, is to seed galaxies above a mass threshold with a BH of mass close to the mass resolution (typically BH seeds of $10^{4-5} \Msun$ within halos of $10^{10-11}\Msun$). This approach leads to a population of BHs at $z=0$ that matches observations \citep[e.g.~][for a recent review]{Kormendy2013}, however it does not shed light on the abundance and properties of faint/small mass BHs at higher $z$, a population which can be important during the initial stages of the EoR.

In this work we attempt a novel approach to model a high-$z$ population of small BHs and study their impact on the EoR. This is done by training a neural network with the properties of the BHs and host galaxies modelled in the cosmological hydrodynamical simulation MassiveBlack-II \citep[\MBII,][]{Khandai2015}. The network is then used to mock the BH population (down to halo and BH masses lower than what was assumed and seeded in \MBII) at redshifts relevant for the EoR. The paper is structured as follows: in section 2 we introduce the simulations and methods employed to develop the neural network; in section 3 we present our results in terms of BH and galactic properties, as well as the impact on the EoR; in section 4 we discuss some caveats and advantages of our new approach and give our conclusions.

\section{Methods}
\label{sec:methods}

In the following we will introduce the simulations and neural network adopted in our work. 

\subsection{Cosmological Simulations}
\label{sec:cosmo_simulation}

We use the cosmological SPH simulation MassiveBlack-II \citep[\MBII,][]{Khandai2015} which tracks baryons, black holes and dark matter down to a redshift of $z=0.0625$ with a dark matter, gas and BH mass resolution of $m_{\rm DM} = 1.1 \times 10^7 h^{-1} {\rm M}_\odot$, $m_{\rm gas} = 2.2 \times 10^6 h^{-1} {\rm M}_\odot$ and $m_{\rm BH, seed} = 5 \times 10^5 h^{-1} {\rm M}_\odot$, respectively. 
BHs are seeded into halos of masses $M_h \geq 5 \times 10^{10} h^{-1} {\rm M}_\odot$. The simulation was run in a box of $100 h^{-1}$ comoving Mpc length, with a $\Lambda$CDM-cosmology with $h=0.701$,  $\Omega_{\rm m} = 0.275$, $\Omega_\Lambda = 0.725$, $\Omega_{\rm b} = 0.046$, $\sigma_8 = 0.816$ and $n_s = 0.968$, where the cosmological parameters have their usual meaning. 

\MBII was post-processed with the multifrequency ionizing radiative transfer (RT) code \texttt{CRASH} \citep{Ciardi2001,Maselli2009,Graziani2013,Graziani2018} between $z=18$ and $z=6$ \citep[hereafter Eide2018 and Eide2020, respectively;][]{Eide2018, Eide2020}, to study the impact of various ionizing and heating sources on the physical properties of the IGM during the EoR. To this aim, stars, high and low mass X-ray binaries (XRBs), shock-heated interstellar gas (ISM) and black holes were identified in \MBII and assigned spectra depending on their physical characteristics as mass, accretion rates, ages, metallicities or local star formation. The corresponding ionizing emissivities were evaluated and used as input for the radiative transfer calculations.

In this work we make use of the cosmological environment and galactic properties provided by \MBII and use them to train a neural network to generate the mass and accretion rate of BHs hosted by such galaxies. In a companion paper we plan, instead, to use numerical simulations as those discussed in Eide2018 and Eide2020 to investigate more in detail the possible impact of the neural network generated BHs on the reionization process of hydrogen and helium.

\subsection{Cosmological, Galactic and BH Properties}
\label{sec:gal_coordinates}
Here we present the 23 galactic and cosmological properties that we use as input to our neural network. 
From \MBII we retrieve
the stellar mass $M_*$ (in $10^{10} h^{-1} {\rm M}_\odot$), 
the mean stellar metallicity ${\rm Z}$, 
the star formation rate ${\rm SFR}$ (in ${\rm M}_\odot\, {\rm yr^{-1}}$), 
the mean stellar age $\tau$ (in ${\rm yr}$), 
the dark matter halo mass $M_h$ (in $10^{10} h^{-1} {\rm M}_\odot$), 
and the galactic gas mass $M_{\rm gas}$ (in $10^{10} h^{-1} {\rm M}_\odot$).
We also derive some geometrical and kinematic properties of the galaxies by doing a principal component analysis of the velocities and positions of the gas and stellar particles \citep[see e.g.~][]{Vanderplas2012}. We find
the galactic gas number density $n_{\rm gas}$ (in ${\rm cm}^{-3}$), 
the mean velocity of the gas
$\{\mu_r^{\rm gas}, \mu_\theta^{\rm gas}, \mu_\phi^{\rm gas}\}$
and of the stars $\{\mu_r^*, \mu_\theta^*, \mu_\phi^*\}$, and their respective velocity dispersion
$\{\sigma_r^{\rm gas}, \sigma_\theta^{\rm gas}, \sigma_\phi^{\rm gas}\}$ and 
$\{\sigma_r^*, \sigma_\theta^*, \sigma_\phi^*\}$,
all in ${\rm km \, s^{-1}}$. 
For each galaxy we also have
the stellar AB luminosity $L_{\rm AB}$ (in ${\rm erg \, s^{-1} \, Hz^{-1}}$),
and the stellar ionizing emissivity $\varepsilon$ (phots s$^{-1}$), 
as calculated in Eide2018.

Additionally, we consider some cosmological properties at the site of each galaxy. Using the cosmic gas number density $n$, we calculate and grid onto $1024^3$ regularly spaced cells the overdensity $\delta = n/\bar{n}$, where $\bar{n}$ is the volume averaged number density.   
As \cite{DiMatteo2017} found that the tidal field plays a central role in the growth of BHs, we follow their prescription to calculate and grid it. We evaluate the strain tensor in Fourier space, $\hat{S} = k^2\hat{\delta}/(k_i k_j)$ from the Fourier transform of the aforementioned gridded overdensity field $\delta$ \citep[following][]{Dalal2008}, and find the tidal field as $T_{ij} = S_{ij} - {\rm Tr}\,S/3$. We calculate the eigenvalues of the tidal tensor, and retain the largest one, $t_1$. As we did for the overdensity, we read off $t_1$ from the grid at the site of the galaxy.

We additionally need to evaluate the accretion rate, luminosity and ionizing emissivity of the BHs. For this, we follow the approach taken in Eide2018 and Eide2020.
In line with \cite{Shakura1973} and the feedback model employed in \MBII, we write the bolometric luminosity as $L = \eta \dot{M}_{\rm BH} c^2$ (in erg s$^{-1}$), where $\dot{M}_{\rm BH}$ is the BH accretion rate, $\eta = 0.1$ is an efficiency parameter and $c$ is the speed of light. The ionizing emissivity $\epsilon_{\rm BH}$ (in phots s$^{-1}$) is derived by rescaling the integrated ionizing spectrum with the bolometric luminosity. The spectrum is determined observationally by \cite{Krawczyk2013} and it is essentially a broken power law at hydrogen-ionizing frequencies, with $L(\nu) \propto \nu^{\alpha}$ and $\alpha = -1$ for $h_{\rm P} \nu > 0.2\,{\rm keV}$, where $\nu$ is the frequency and $h_{\rm P}$ is the Planck constant. The integral of the ionizing spectrum gives the emissivity. From the rescaled spectrum we also derive the AB luminosity of the BHs, $L_{\rm AB, BH}$ (in erg s$^{-1}$ Hz$^{-1}$). 

\subsection{The Neural Network}
\label{sec:bhs}
We now describe how we construct and train the neural network which ultimately is used to predict the BH masses, $M_{\rm BH}$, and accretion rates, $\dot{M}_{\rm BH}$. In essence, these are derived from the aforementioned $23$ galactic properties, and the network is trained and validated on existing BHs at $z=6$.

We use \texttt{TensorFlow}\footnote{\url{https://www.tensorflow.org}} with the \texttt{Keras}\footnote{\url{https://keras.io}} interface to construct the neural network, while we employ \texttt{RMSProp} as optimiser\footnote{For an overview of gradient descent optimization algorithms (including \texttt{RMSProp}) we refer the reader to https://ruder.io/optimizing-gradient-descent/index.html}. Starting from a single input layer $L(23)$ with the same number of units as we have learning parameters ($23$), we consecutively add layers with larger number of units to the network and test its accuracy after adding each new layer.
We eventually arrive at a multilayered deep network where introduction of additional hidden layers lead to overfitting and modelling of the noise in the data because of the too many free parameters. We then introduce dropout layers, $D(r,v)$, which when enabled ($v=1$) randomly remove a fraction $r$ of the connections to the preceding layer, helping to increase the versatility of the network and to prevent overfitting \citep{Hinton2012}. The maximally connected network $f$ can be described as one that takes an input vector $\mathbf{x}$ of our $23$ learning (and prediction) parameters and forward feeds it through several hidden layers $L$ and $D$ before finally reaching an output layer which returns the predictions $\mathbf{y}^{\rm NN}$, where $y_0^{\rm NN} = M_{\rm BH}^{\rm NN}$ and $y_1^{\rm NN} = \dot{M}_{\rm BH}^{\rm NN}$. In its most complex form, it has the following structure,
\begin{align}
    \mathbf{y}^{\rm NN} &= f\left(\mathbf{x}; (LD(0.5,1))_3 L_3\right) = (LD)_3 L_3 \nonumber \\
    &= \mathbf{x} \to L(23) \nonumber \\ 
    &\hphantom{= \mathbf{x}\; } \to D(0.5,1) L(92) \nonumber \\
    &\hphantom{= \mathbf{x}\; } \to D(0.5,1) L(8)  \nonumber \\
    &\hphantom{= \mathbf{x}\; } \to D(0.5,1) L(16) \nonumber \\ 
    &\hphantom{= \mathbf{x}\; } \to L(16) \nonumber \\
    &\hphantom{= \mathbf{x}\; } \to L(2) 
    \label{eq:network}
\end{align}
where the arrows indicate that the outputs $\mathbf{a}^{l-1}$ of the layer $l-1$ are used to compute the activation of the units in the next hidden layer $l$, $\mathbf{a}^{l} = {\tt ReLU}((1-r)^{-1}{\rm D}^l{\rm W}^l \mathbf{a}^{l-1} + \mathbf{b}^l)$. Here, ${\rm W}$ is the matrix of elements $W^l_{ki}$ of the connection weights between unit $i$ of layer $l-1$ and unit $k$ of layer $l$, $D^l(r,1) = (1-r)^{-1}{\rm D}^l$ is a matrix where a fraction $r$ of the connections are dropped, $\mathbf{b}$ is a bias, and \texttt{ReLU} is the activation function. The final layer has a linear activation function, i.e.~$\mathbf{y} = \mathbf{a}^{N} = {\rm W}^N \mathbf{a}^{N-1} + \mathbf{b}^N$. In the following we omit `$r$' from the notation as we always assume the standard vale $r=0.5$ \citep{Hinton2012}. 

\begin{figure}
\centering
\includegraphics[width=1.0\columnwidth]{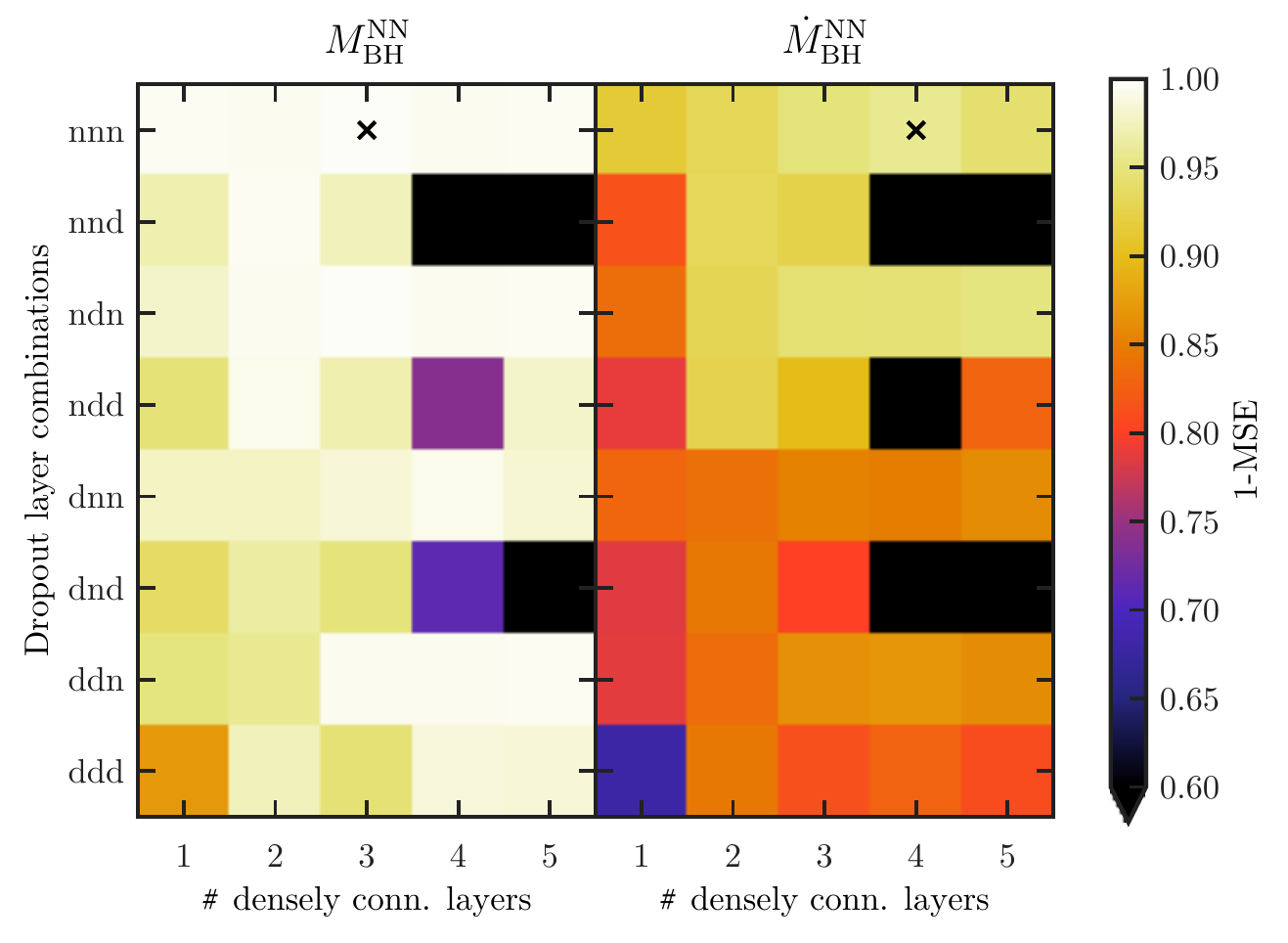}
\caption{Impact of network composition on the accuracy 1-MSE in predicting BH masses $M^{\rm NN}_{\rm BH}$ (left panel), and accretion rates $\dot{M}^{\rm NN}_{\rm BH}$ (right panel). The $x$-axis shows the number of densely connected layers in the network, while the $y$-axis shows the combinations of dropout layers. The crosses indicate the networks with the highest accuracy.}
\label{fig:bh_network_composition}
\end{figure}

In Fig.~\ref{fig:bh_network_composition} we show how the combinations of the hidden layers affect the accuracy 1-MSE (mean-square error) of the $M$ predictions, with ${\rm MSE} = M^{-1} \sum^M_{p=1} (y_{i,p}^{\rm NN} - y_{i,p})^2$ for $i=0,1$ and where $y_i$ are the validation values. We test the network on the portion of the dataset that it has not been trained on. The figure shows combinations of $N=1\dots5$ hidden layers in addition to the output layer. A network where $N=1$ only has two \texttt{ReLU} layers, $LD(v)_3L$. We also show combinations of the dropout layers $D$. Networks where all the dropout layers are enabled are labeled as `ddd' in the figure, whereas a combination such as e.g.~$LD(1)(LD(0))_2 L_3 = LDL_4$ or $LD(1)D(0)_2L = LDL$ is labeled `dnn'. 

The networks without any dropout layers, labeled `nnn', usually have the smallest errors, but also the largest potential for overfitting the data. The networks with the smallest error (marked in the figure with a hatch) are $L_4$ for the BH mass, and $L_5$ for the BH accretion rate. The `d**' networks, where a dropout is applied right after the input layer, generally present larger errors, particularly for the predicted BH accretion rates. Unsurprisingly, the predictions are better with dropout layers for the $N>1$ layered networks. The `ndn' class of networks particularly sets itself apart with consistently good predictions. We find that the $LDL_2$ network is the one that strikes the best balance between simplicity and predictive power, and is the one we apply in our work.

The training of all the networks was done at $z=6$, where \MBII has a sizable population of $2,734$ BHs. 
As the distribution of accretion rates is not uniform in our sample of galaxies hosting BHs, we whitened the input data before training. The whitening is done by duplicating galaxies with rare accretion rates, where the properties of the duplicates are added gaussian noise $\sim \mathcal{N}\left(0, 0.05 \boldsymbol{\sigma}\right)$ based on the variance $\boldsymbol{\sigma}$ of these properties.  This extends the training sample and prevents the network from being biased towards only predicting the most common BH masses and accretion rates.

Furthermore, we did not train the networks on galaxies holding BHs with masses equal to those of the seeds, but rather restricted ourselves to $M_{\rm BH} > 1.1 m_{\rm BH, seed}$, as the BH properties just after seeding do not immediately reflect the properties of the host galaxy. This left $62\%$ of the data available for training and verification. This also means that any prediction in the range $1 \leq M_{\rm BH}/m_{\rm BH, seed} \leq 1.1$ can be used to evaluate the predictive power of the network for masses that it has not been trained for.

We used $7/8$ of the full sample of galaxies in the whitened set at $z=6$ for training the network, before validating it on the remaining $1/8$ of the set. By evaluating the MSE of the predictions of the network versus the validation data, we conclude that our ability to predict the BH masses and accretion rates with our chosen $LDL_2$ network happens with an accuracy 1-MSE of $0.995$ and $0.936$, respectively.

\section{Results}
\label{sec:results}

In this section we present our results in terms of galactic and BH properties, as well as the impact that our network predicted BH population has on the reionization process. 

\begin{figure}
\centering
\includegraphics[width=1.0\columnwidth]{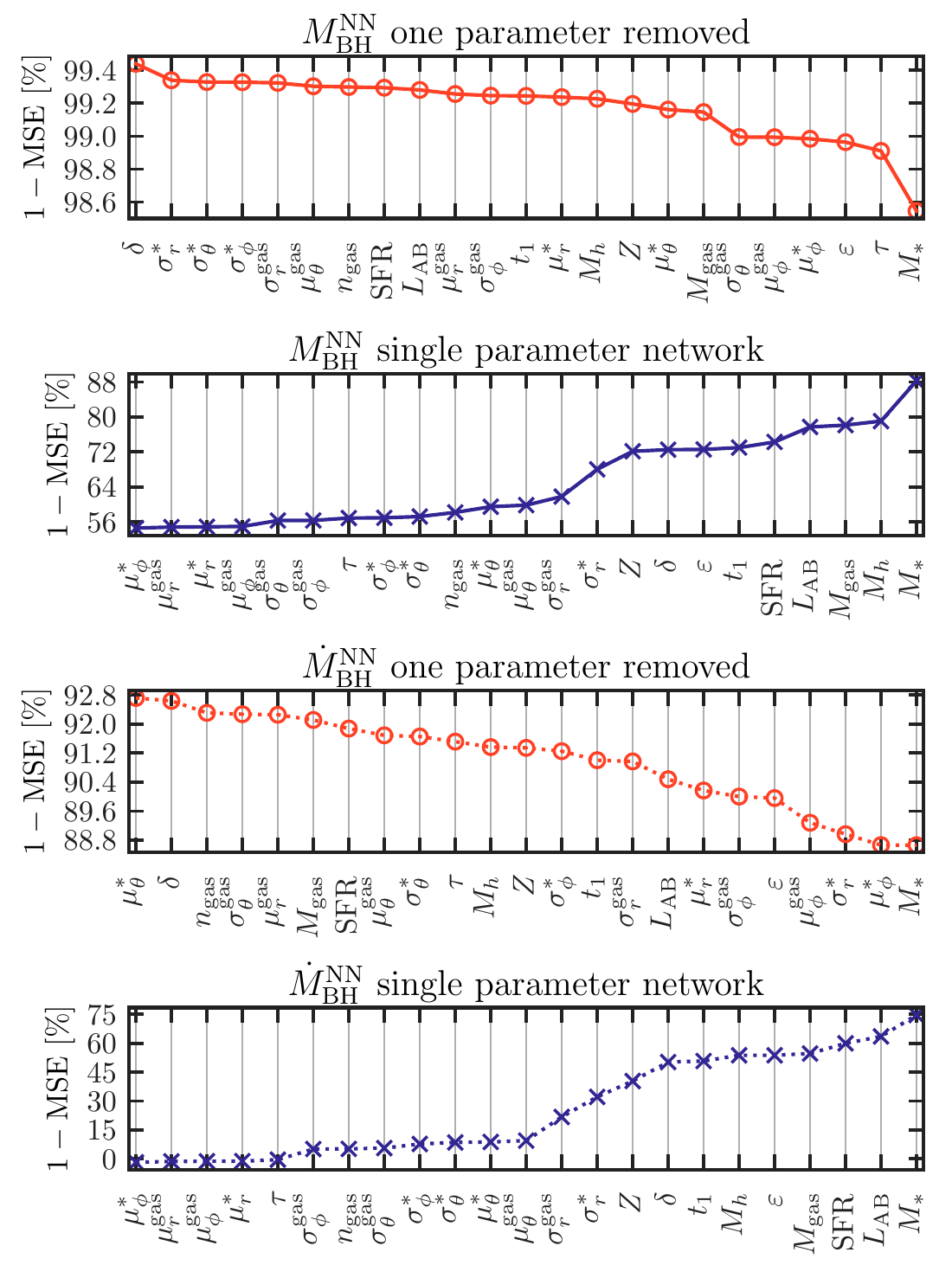}
\caption{Network accuracy 1-MSE in predicting BH masses $M^{\rm NN}_{\rm BH}$ (solid lines), and accretion rates $\dot{M}^{\rm NN}_{\rm BH}$ (dotted lines). The red circled lines refer to the accuracy of the network in the absence of the parameter indicated in the x-axis, while the blue crossed lines refer to networks with only the removed parameter. 
}
\label{fig:bhseed_importance}
\end{figure}

\begin{figure}
\includegraphics[width=\columnwidth]{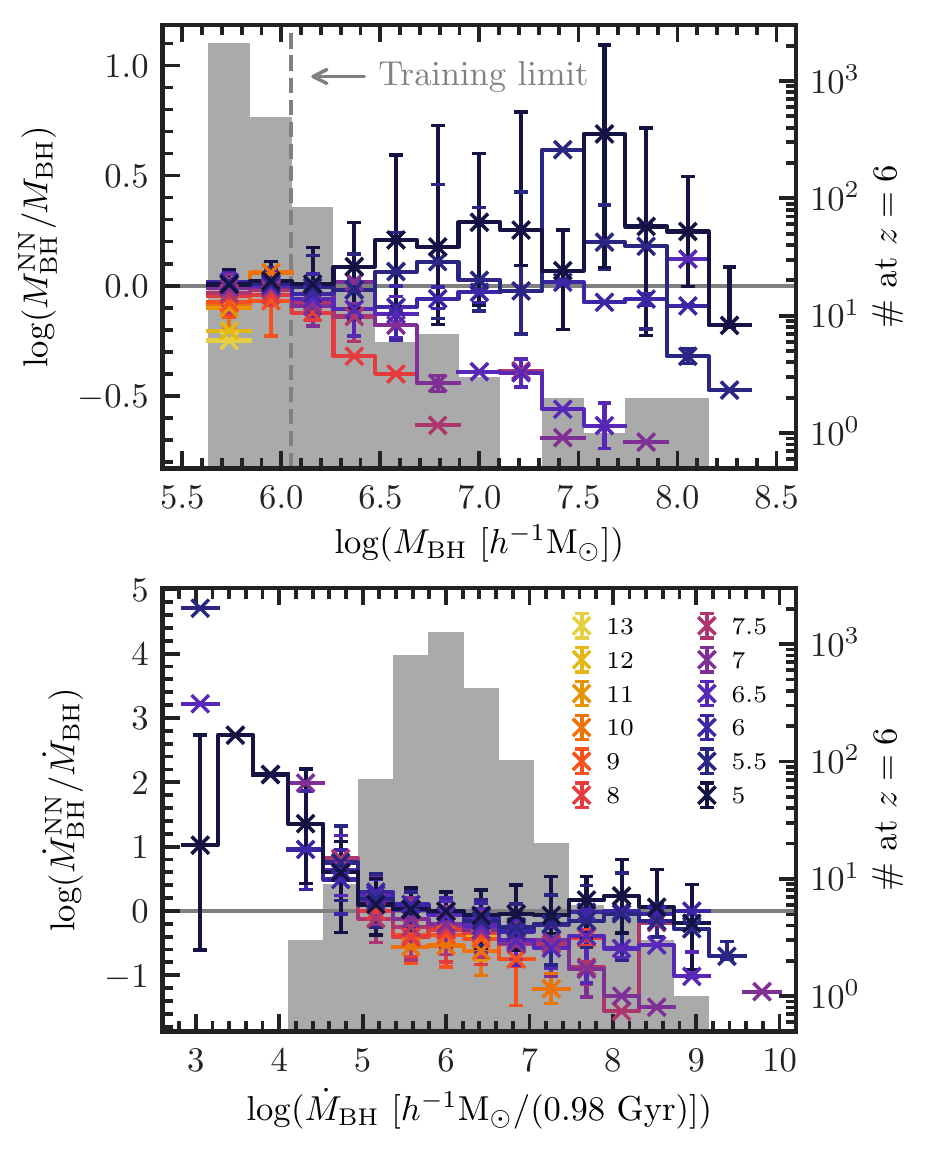}
\caption{Deviation between the predictions from the neural network trained at $z=6$ and the actual values, for galaxies harbouring BHs between $z=13$ and $z=5$, indicated by the line colour. The crosses, lower and upper limits give the median, $16$th and $84$th percentiles, respectively, of the associated bins. We show histograms of the masses and accretion rates of the \MBII BHs at $z=6$ in the background. The vertical dashed line indicates the lower training limit on the mass, $1.1 m_{\rm BH, seed}$.  \textit{Upper panel:} predicted, $M_{\rm BH}^{\rm NN}$, against true, $M_{\rm BH}$, BH mass. \textit{Lower panel:} predicted, $\dot{M}_{\rm BH}^{\rm NN}$, against true, $\dot{M}_{\rm BH}$,  BH accretion rate.
}
\label{fig:tensorflow_match}
\end{figure}

\subsection{Relation between Galactic and Black Holes Properties}
\label{sec:results_main_comps}
We now turn to examine if any of the 23 galactic properties plays a dominant role in predicting the BH masses and accretion rates.  We do this by generating (i) a network which is the same as the original one except that now one parameter is removed, and (ii) a network using solely this parameter. In both cases we estimate the MSE on the predicted BH masses and accretion rates.

In Fig.~\ref{fig:bhseed_importance} we show the networks' accuracy, 1-MSE, in predicting BH masses and accretion rates. A low $1-{\rm MSE}$  for the models plotted in red  reflects a poorer performance of the network without the component under consideration. Conversely, a high $1-{\rm MSE}$ for the models in blue means that the predictive power of this single-parameter network is better. 

We first note that the multi-parameter networks have a higher accuracy ($>88$\%) compared to the single-parameter networks ($< 88$\%), and are hence performing better. The best single parameter for predicting the BH accretion rate and mass is $M_*$, for which the networks recover $M_{\rm BH}$ and $\dot{M}_{\rm BH}$ with an accuracy of $88$\% and $75$\%, respectively. As for the multi-parameter networks, they perform worst when removing $M_*$, yielding an accuracy of $99$\% and $89$\% for $M_{\rm BH}$ and $\dot{M}_{\rm BH}$, respectively. 

While the relevance of $M_*$ is clear both for single- and multi-parameter networks, this is not the case for the other parameters. The three next-most important parameters for the determination of $M_{\rm BH}$ are the mean stellar age $\tau$, the stellar ionizing emissivity $\varepsilon$ and the $\phi$ component of the mean velocity of the stars $\mu_\phi^*$ for the multi-parameter network. For the single-parameter network, instead, these are the dark matter halo mass $M_h$, the galactic gas mass $M_{\rm gas}$ and the stellar AB luminosity $L_{\rm AB}$. Similarly, for $\dot{M}_{\rm BH}$ the most relevant quantities in the multi-parameter network are the $\phi$ component of the mean velocity of the stars $\mu_\phi^*$, the $r$ component of the stellar velocity dispersion $\sigma_r^*$ and the $\phi$ component of the mean velocity of the gas $\mu_\phi^{\rm gas}$, indicating that the network captures the dependency between accretion and environmental kinematics; while $L_{\rm AB}$, SFR and $M_{\rm gas}$ yield the highest accuracies in the single-parameter networks. We recover the same order of importance for the single-parameter networks by calculating the correlation coefficient between $M_{\rm BH}$ or $\dot{M}_{\rm BH}$ and the parameter in question. Again, it should be noted that the predictions of these single-parameter networks are far less accurate ($\sim 60$\%) than the multi-parameter networks ($> 89$\%).
As a further test, we create a network with only $M_*$, $L_{\rm AB}$, $M_{\rm gas}$ and $M_h$ as input parameters. It recovers $M_{\rm BH}$ and $\dot{M}_{\rm BH}$ with an accuracy of $97$\% and $85$\%, respectively. This highlights our need for the full network's complexity if the goal is to recover the accretion rates as precisely as possible.

\subsection{Populating Galaxies with Black Holes}
\label{sec:results_populating_all}

Our network was generated from $z=6$ BHs and their host galaxies. We now turn to examine how it performs at other redshifts, for galaxies it has not been trained for. This can also reveal any redshift evolution in BH properties.

To do so, we use the network to seed BH-hosting galaxies at various $z$ with BHs with predicted masses $M_{\rm BH}^{\rm NN}$ and accretion rates $\dot{M}_{\rm BH}^{\rm NN}$, and compare them to the  $M_{\rm BH}$ and $\dot{M}_{\rm BH}$ directly obtained from \MBII. 
We show the deviation between the generated and true values at various redshifts in Fig.~\ref{fig:tensorflow_match}. In the $m_{\rm BH, seed} < M_{\rm BH} < 10^{6.5} h^{-1} {\rm M}_\odot$ mass range, the deviations vary from $20\%$ larger to $30$\% smaller, with the largest ones at $M_{\rm BH} > 10^7 h^{-1} {\rm M}_\odot$, where we have a poorer statistic of the training set. At $M_{\rm BH} < 1.1 m_{\rm BH, seed}$ where the network has not been trained for (this mass limit is indicated by a vertical dashed line in the figure), the predictions are $2$--$4$\% larger than the true values at $z=6$, while at $z=9$ they are $\sim 15\%$ lower. This indicates that our network is very powerful in predicting masses it has not been trained for. The predicted accretion rates deviates from being between $\sim 30$\% larger to $\sim 60$\% smaller for $10^5 < \dot{M}_{\rm BH}/(h^{-1} {\rm M}_\odot / (0.98\, {\rm Gyr})) < 10^{7.5}$. Also in this case, the predictions are best within the most common range of accretion rates. At the high mass and accretion rate end, the network underpredicts the true values at $z > 6$ and overpredicts them at $z < 6$. This indicates that the BH formation efficiency declines with decreasing $z$. We also see this effect within the central mass and accretion ranges, albeit in a much more moderate fashion---e.g.~at $z=9$ ($z=5$), $M_{\rm BH} = 10^6 h^{-1} {\rm M}_\odot$ BHs are on average predicted to be $16\%$ less ($4\%$ more) massive, and the accretion rate of $\dot{M}_{\rm BH} = 10^6 h^{-1} {\rm M}_\odot/(0.98\,{\rm Gyr})$ BHs is predicted to be $\sim 50\%$ ($\sim 4\%$) lower---indicating that this is not merely an effect caused by lacking statistics of our training set.

\begin{figure}
\centering
\includegraphics[width=\columnwidth]{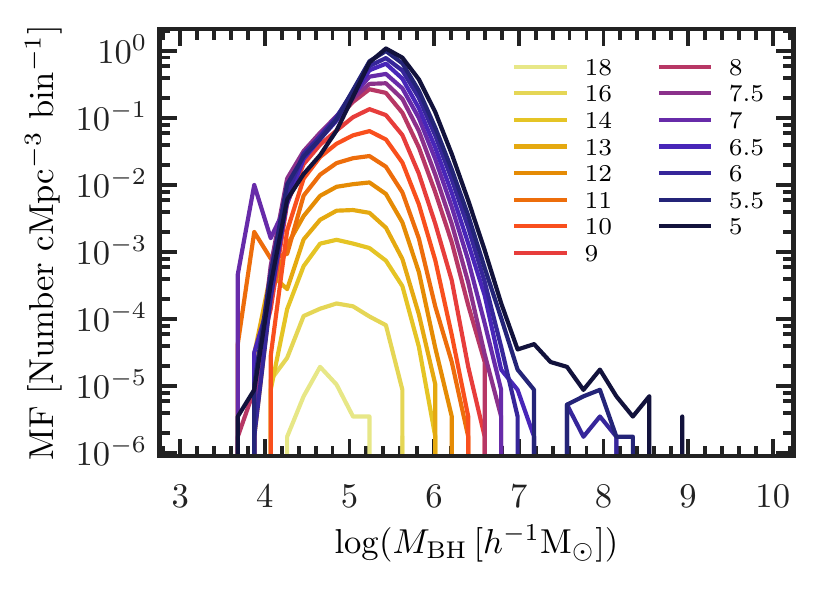}
\caption{Mass function of BHs seeded with our neural network in every galaxy. The line color indicates different redshifts, in the range $z=5-18$.}
\label{fig:BHMF}
\end{figure}

Next, we populate all galaxies in the range $z=5-18$ with a BH using the neural network including all the 23 physical properties described in sections \ref{sec:gal_coordinates} and \ref{sec:bhs}. We thereby create a much larger population of BHs than is present in \MBII. 
In Fig.~\ref{fig:BHMF} we show the resulting mass function at various redshifts. While at $z=18$ the BH population is limited to the range $10^4 < M_{\rm BH}/(h^{-1} {\rm M}_\odot) < 10^5$, the peak of the mass function shifts towards higher values with decreasing redshift, and by $z \sim 6$ we have BHs with masses as high as $\sim 10^8 h^{-1}$~M$_\odot$. The smallest BHs have $M_{\rm BH} \sim 10^{3.6} h^{-1}$~M$_\odot$ at all times. This is more than a magnitude lower than the seed mass $m_{\rm BH, seed}$ of \MBII, and reflects that the predictions of the network are not restricted by the mass range it was trained on.
Note that the generated BH mass function is not dissimilar to those for a range of physical BH seed models at $z=15-18$ \citep[e.g.;][]{Volonteri2008}. Our generated BH population
appears to exploit reasonably well the actual resolution of the simulation, introducing BHs at smaller masses and earlier time when they are indeed expected to form.
We note here that seeding halos of mass smaller than the one used in the MBII prescription is not a mere extrapolation, but is made possible by the fact that, even if the mass falls outside of the range used for the training, all the other 22 properties are not restricted by any limit. Hence the robustness of our procedure.

\begin{figure}
\includegraphics[width=\columnwidth]{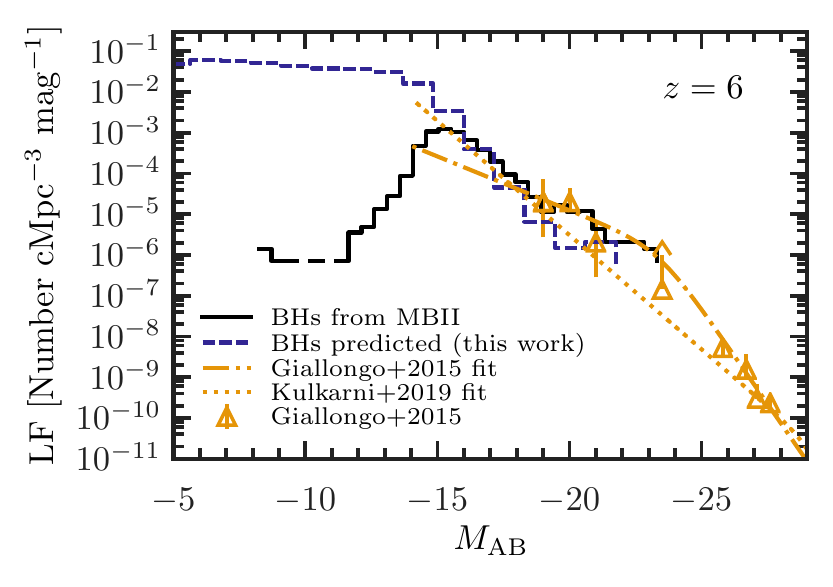}
\caption{Luminosity functions at $z=6$. The lines refer to the LF of the BH population from \MBII (solid black line), the LF of the BHs seeded with our neural network (dashed blue line), and observational constraints from \protect\cite{Giallongo2015} (data: yellow triangles, fit: dashed-dotted line) and \protect\cite{Kulkarni2019} (fit: yellow dotted line) .}
\label{fig:QLF}
\end{figure}

In Fig.~\ref{fig:QLF} we show the UV luminosity function (LF) of the BHs at $z=6$, and compare it to the LF of the \MBII BHs, as well as to the observationally determined LFs of \cite{Giallongo2015} and \cite{Kulkarni2019}. 
As our network slightly underpredicts the highest accretion rates, we have a small deficit of bright BHs compared to both the \MBII-seeded BHs and the \cite{Giallongo2015} observations. It should be noted, though, that the bright end of the observed LF may be overestimated \citep{Parsa2018}, so that our conservative result might be more realistic. This is further corroborated by the recent compilation of \cite{Kulkarni2019}, based on 66 QSOs at $5.5 < z < 6.5$, as our predicted LF matches their observations at all $M_{\rm AB}$.
The agreement of our LF with the original one from the \MBII and the \cite{Giallongo2015} LF is extremely good in the range $-17 < M_{\rm AB} < -15$. Our network also predicts a substantial population of faint BHs which are not present in \MBII, and yields a LF with a knee at $M_{\rm AB} = -15$ and no turnover at least down to $M_{\rm AB} = -5$.

\begin{figure}
\includegraphics[width=\columnwidth]{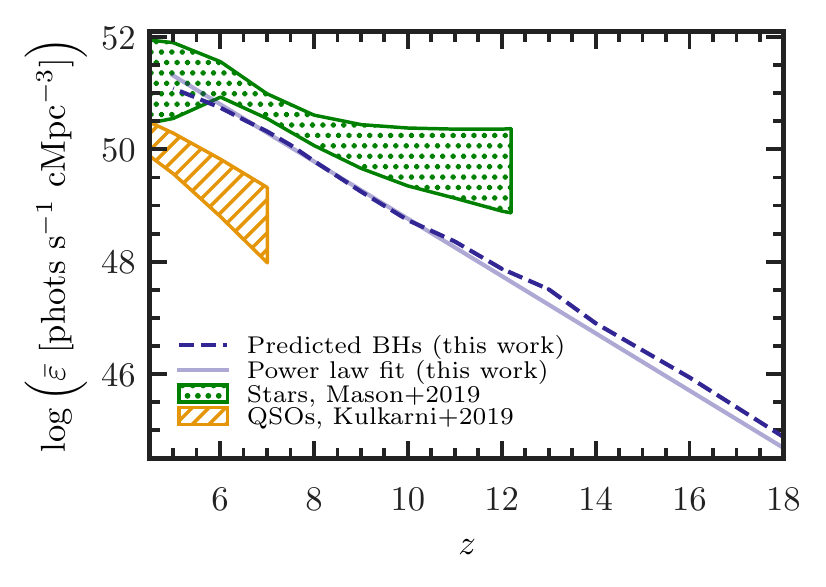}
\caption{Volume averaged hydrogen ionizing emissivity $\bar{\varepsilon}$ (in phots s$^{-1}$ cMpc$^{-1}$) from our predicted BHs (dashed blue line), with a power law fit (solid pale blue line). The hatched yellow area refers to the constraints derived by \citet{Mason2019} based on observed bright QSOs compiled by \citet{Kulkarni2019}. The upper (lower) limit includes QSOs with $M_{\rm AB}<-14$ (-21). The dotted green areas are inferences of the stellar contributions from observational constraints by \protect\citet{Mason2019}.
}
\label{fig:ems_predicted}
\end{figure}

In Fig.~\ref{fig:ems_predicted} we plot the comoving volume averaged emissivity, $\bar{\varepsilon}$, in comparison to values inferred from observations. The predicted emissivity increases exponentially from $z=18$, when $\bar{\varepsilon} = 7.6 \times 10^{41}$ phots s$^{-1}$ cMpc$^{-1}$, to $z=5$, where $\bar{\varepsilon} = 1.2 \times 10^{52}$ phots s$^{-1}$ cMpc$^{-1}$. This evolution can be parametrized as  a power law,
\begin{equation}
    \log \bar{\varepsilon}(z) = -0.5097 z + 53.86,
    \label{eq:power_fit_ems}
\end{equation}
using a least-square fit to the predictions.
We find that the predicted emissivity is much higher than that inferred by \cite{Mason2019} based on the \cite{Kulkarni2019} sample of bright QSOs (yellow hatched area), suggesting that the contribution to the ionizing budget of our faint population is significant, although it should be noted that this is an upper limit as we have populated every galaxy with a BH. Our predicted $\bar{\varepsilon}$ is however below the inferred contributions from stars, as shown in the non-parametric model inferred by \cite{Mason2019} from the CMB optical depth, dark $\Lya$ and ${\rm Ly}\beta$ pixels and hydrogen neutral fraction constraints from $\Lya$ observations. Our $\bar{\varepsilon}$ overlaps with the \cite{Mason2019} model at $z \lesssim 6$.

\subsection{Impact on the Reionization Process}
\label{sec:results_eor}

\begin{figure}
\includegraphics[width=\columnwidth]{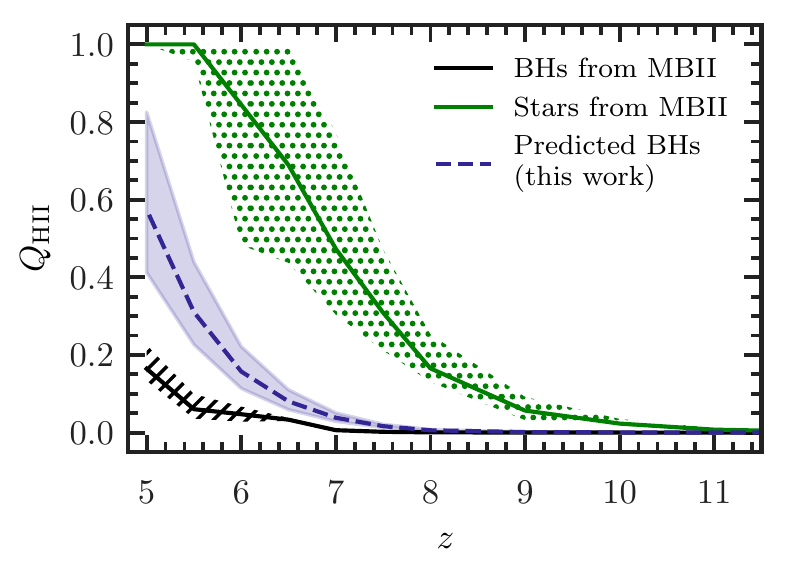}
\caption{\HII volume filling factor $Q_{\rm HII}$ as a function of redshift $z$. The lines refer to a case in which the reionization process is driven by \MBII seeded BHs (solid black line in hatched area), stars in \MBII (green line in dotted area) and BHs populated by our neural network (dashed blue line). The upper and lower limits refer to clumping factors $C=1$ and $C=10$, respectively.
}
\label{fig:Q_volfilling}
\end{figure}

The final question we address in our study is whether such a population of faint BHs could have a significant impact on the EoR. While we plan to run simulations as those presented in Eide2018 and Eide2020 including these faint BHs, here we limit the analysis to a simpler approach. We calculate the filling factor $Q_{\rm HII}$ of ionized hydrogen (\HII) as (\citealt{Madau1999}):
\begin{equation}
	\frac{dQ_{\rm HII}}{dt} = \frac{f_{\rm esc}\bar{\varepsilon}}{\bar{n}_{\rm H}} - \frac{Q_{\rm HII}}{\bar{t}_{\rm rec}},
	\label{eq:Q_volfilling}
\end{equation}
where $f_{\rm esc}$ is the escape fraction of ionizing photons, $\bar{\varepsilon}$ is the volume averaged ionizing emissivity, $\bar{n}_{\rm H}$ is the average cosmic hydrogen number density and $\bar{t}_{\rm rec} = (C \bar{n}_{\rm H} \alpha(T))^{-1}$ is the recombination time, for which we assume a clumping factor $C=1, 5, 10$ and a case-A recombination coefficient $\alpha$ at $T=10^4\,{\rm K}$. 
We calculate $Q_{\rm HII}$ for the \MBII BHs, as well as for those seeded by our neural network, assuming $f_{\rm esc} = 1$ for both. As a comparison, we also calculate $Q_{\rm HII}$ for the stars of \MBII, assuming $f_{\rm esc} = 0.15$ as in Eide2018 and Eide2020.
These are shown in Fig.~\ref{fig:Q_volfilling}.
For $C=5$ we find that the population of mainly faint BHs seeded with our neural network results in a reionization history in which the BHs have a central, albeit not dominant, role, reaching $Q_{\rm HII} > 0.15$ (0.5) at $z = 6$ (5). This is in stark contrast to the massive BHs of \MBII, which reside only in the most massive galaxies and yield $Q_{\rm HII} < 0.05$ (0.2) at the same redshifts. As expected, the stars dominate the reionization process, producing $Q_{\rm HII} \sim 1$ already at $z=6$.
Finally, we should note again that the contribution from the network generated BHs should be regarded as an upper limit, as not every galaxy is in reality expected to host
 an active BH. We defer to future work a refinement of this approach.

\section{Discussion and conclusions}
\label{sec:conclusions}
In the cosmological hydrodynamical simulation MassiveBlack-II \citep{Khandai2015}, galaxies with a halo mass in excess of $M_{h, \rm seed} = 5 \times 10^{10} h^{-1}~{\rm M}_\odot$ are populated with seed black holes (BHs) with $m_{\rm BH, seed} = 5.5 \times 10^5h^{-1}~{\rm M}_\odot$.
While this prescription assures that the BH population has physical properties consistent with observations at $z \lesssim 6$, a different seeding procedure, with BHs hosted also in smaller galaxies, might have a strong impact, among others, on the role played by BHs in the reionization process of the intergalactic medium and the related 21~cm signal.
To investigate this in more detail, we have trained a neural network using the properties of galaxies harboring BHs at $z=6$.
This network allowed us to mock BHs in all galaxies down to the resolution limit of the simulations at all redshifts, corresponding to halos of mass $\sim 9 \times 10^6 h^{-1}$~M$_\odot$.
By design and through training, the network replicates the properties of the pre-existing BHs in the simulation.

Our network predicts the BH masses and accretion rates of existing BHs with great precision ($>99\%$ and $>93\%$, respectively). Interestingly, we mock BHs with masses below the \MBII seed mass when applying the network to all galaxies, also those with halo masses below $M_{h,\rm seed}$. Although the seeding procedure is extrapolated to lower masses, our predictions of $M_{\rm BH}$ and $\dot{M}_{\rm BH}$ are robust because they are constrained in $23$ dimensions with a high accuracy (e.g.~the predictions of the mass function in Fig.~\ref{fig:BHMF} where BHs are lighter at higher $z$). In fact, a galaxy with $M_h < M_{h, \rm seed}$ may still share up to $22$ other parameters with galaxies hosting BHs in \MBII, and thus be tightly constrained in these other dimensions. Additionally, as the networks have been trained on galaxies with $M_{\rm BH}>1.1 m_{\rm BH,seed}$, the predictions in the range $1 \leq M_{\rm BH} \leq m_{\rm BH,seed}$ have been used to confirm the strong predictive power of the networks for masses that they had not been trained for.

We find that removal of one parameter, including $M_h$, from our network did not lead to a significant deterioration of its predictions. Similarly, not a single one of the input parameters provides predictions as accurate as the full network.
The exercise of removing parameters from the network, nevertheless, highlighted that the stellar mass of the galaxy, $M_*$, is the most important parameter. Alone, it can predict the BH mass with an accuracy of $0.88$, while $M_h$ has an accuracy of $\sim 0.80$. It is harder to infer the effect of the velocity dispersion. From the well-known $M_{\rm BH}$--$\sigma^*$ relation \citep{Ferrarese2000} we expect the velocity dispersion to be a dominant parameter, but we cannot directly infer its role as it is not a single input to our network, but it is rather decomposed along each coordinate axis $r$, $\phi$ and $\theta$. 

Even though the formation efficiency of BHs is declining with decreasing redshift, this does not necessarily imply that their growth is decoupled from the stellar growth. On the contrary, our network has a strong dependence on stellar properties, such as stellar mass and age. Observations indicate that the SFR history is closely related to the BH accretion history \cite[see e.g. the review by~][]{Madau2014}, but they are not identical. Furthermore, our power-law parametrisation of the BH emissivity with a slope of $-0.5$ is similar to that of $-0.45$ which has been found for the stellar UV density at $z>9$ \citep[][]{Oesch2018,Madau2018}, although our BH emissivity relation lacks the turnover to a slower growth at $z\leq 9$, which is seen instead for the stellar UV density. The strong dependence on the tidal field, overdensity, gas mass and halo mass also indicates that an environment that promotes stellar growth also positively influences BH growth. Such highly biased regions are in fact required to avoid quenching of the growth of the lightest BHs by SN-feedback \citep[][]{Inayoshi2019}. 

While we took great care in the training of the network, its performance is still somewhat limited by the size of the training sample, both in terms of number of objects and range of masses covered. Furthermore, our network was trained on $z=6$ galaxies hosting a BH, as only at that time does \MBII produce a sizable population of BHs. This situation would improve by adopting larger and/or higher resolution simulations, such as \textsc{BlueTides} \citep{Feng2015} or Illustris TNG300 \citep{Nelson2018}, or employ simulations specifically designed for this task.
Nevertheless, we found that the network predicts the properties of the majority of BHs at all redshifts with high accuracy, indicating also that there is no significant evolution in the relation between the environment and the BHs' properties, in line with \cite{Huang2018}. However, our slight deficiency of brighter BHs at $z>6$ (and surplus at $z<6$) points to these being formed more efficiently at early times (see also e.g. \citealt{Degraf2012,DeGraf2015}). Our results suggest that a galaxy at $z>6$ with properties identical to those of one at $z=6$ is more likely to host a brighter BH.

We also note that recent work which relaxes repositioning of the BHs (as done instead in \MBII and most large scale cosmological simulations such as the previously mentioned Illustris) and uses additional dynamical friction (e.g. \citealt{Tremmel2018,Pfister2019,Barausse2020}) should provide more realistic predictions for the early BH populations and their BH merger rates, possibly leading to lower occupation fractions and central BH masses in the galaxies. In the future, different scale simulations
(such as those mentioned above) could be used as additional training sets.

Our slight deficiency of the brightest BHs at lower $z$ in turn ensures a perfect match at $z=6$ to the recent LF of \cite{Kulkarni2019}, and a perfect match at $-15 > M_{\rm AB} > -17$ to the LF of \cite{Giallongo2015}. The most interesting feature of our results is however the large population of faint, $M_{\rm AB} > -15$, BHs. Such a population is entirely possible, as the pre-existing BHs (and the combined contributions from other energetic X-ray emitting sources) in \MBII are unable to account for more than a few per cent of the unresolved X-ray background \citep{Ma2018}, leaving ample margin for a higher contribution at high redshift.

This predicted population of BHs is unable to drive EoR alone, but it may play an important role nevertheless. Our mocked BHs do not yield enough ionizing photons to fulfil the constraints on the ionizing budget calculated from observational constraints by \cite{Mason2019}. However, our emissivites are an order of magnitude larger than those inferred from integrating the LF of the brighter QSOs of \cite{Kulkarni2019}. 
Our BHs leave a significant imprint on the \HII volume filling factor, which  at $z=5$ ranges from $Q=0.41$ with a clumping factor $C=10$ to $Q=0.83$ with $C=1$. The existing BHs in \MBII can at best yield $Q=0.23$ with $C=1$, but while this population satisfy the bright end of the LF down to $z>2$, it does not include the fainter population that our network predicts.
Our population of mocked BHs is neither negligible, nor is it as dominating as the one of \cite{Madau2015}. Further work is needed to investigate whether they will induce an extended \HeII reionization epoch as observations imply \citep{Worseck2016,Worseck2019} without providing undue heating \citep[see e.g.~][]{DAloisio2017,Garaldi2019}. We plan to investigate this more in detail with numerical simulations following the work of Eide2018 and Eide2020.

A more prominent population of high-$z$, small mass BHs could also have an important impact on the 21~cm signal from neutral hydrogen in the IGM, by partially ionizing and heating the gas prior to full reionization (e.g. \citealt{Madau1997}).

Our conclusions can be summarized as follows.
\begin{itemize}
\item We train a neural network on properties of BH hosting galaxies at $z=6$. For our training sample, this predicts the mass,  $M_{\rm BH}$, and accretion rate, $\dot{M}_{\rm BH}$ of BHs with an accuracy $>99\%$ and $>93\%$, respectively. These properties at other redshifts are also predicted with high precision. 
\item $M_{\rm BH}$ and $\dot{M}_{\rm BH}$ are predicted with the most relevant single parameter, the stellar mass $M_*$, with an accuracy of $88$\% and $75$\%, respectively. Removing $M_*$ degrades the network to accuracies of $98.6$\% and $88.8$\%. The predictions of our network are robust, even when single parameters are ill-defined.
\item The neural network is slightly less effective at predicting the brightest and most massive BHs at $z>6$, and conversely predicts a population of slightly brighter BHs at $z<6$. This points to a decrease in BH formation efficiency with decreasing $z$.
\item Populating all galaxies with a nuclear BH, we predict a substantial population with mass below that of the seeds at all redshifts. This results in a LF at $z=6$ with a knee at $M_{\rm AB} = -15$ and a lack of turnover at least down to $M_{\rm AB} = -5$.
\item Our predicted population of BHs can contribute significantly to H reionization, yielding a Universe in which H is $\sim 15$\% ionized by BHs at $z=6$ for a clumping factor of 5. The bright BHs alone, which are well reproduced by \MBII, predict instead a Universe that is only $\sim 5\%$ ionized at the same redshift. 
\end{itemize}

\section*{Acknowledgments}
We thank Enrico Garaldi, Martin Glatzle and Max Gr\"onke for enlightening discussions, and an anonymous referee for useful comments.
MBE thanks ITA in Oslo for the hospitality.
TDM acknowledges funding from NSF ACI-1614853,  NSF AST-1616168, NASA ATP 19-ATP19-0084 and NASA ATP 80NSSC20K0519,
ATP 80NSSC18K101.

We have greatly benefited from the availability of open-source software. In this work we have made use of
\texttt{Scikit-Learn} \citep{scikit-learn},
\texttt{Keras} \citep{keras},
\texttt{Tensorflow} \citep{tensorflow},
\texttt{Matplotlib} \citep{Hunter2007},
\texttt{Numpy} \citep{VanderWalt2011} and
\texttt{SciPy} \citep{virtanen2019}
.

\section*{Data availability}

No new data were generated or analysed in support of this research.

\bibliographystyle{mnras} 
\bibliography{references2} 

\begin{thebibliography}{}
\makeatletter
\relax
\def\mn@urlcharsother{\let\do\@makeother \do\$\do\&\do\#\do\^\do\_\do\%\do\~}
\def\mn@doi{\begingroup\mn@urlcharsother \@ifnextchar [ {\mn@doi@}
  {\mn@doi@[]}}
\def\mn@doi@[#1]#2{\def\@tempa{#1}\ifx\@tempa\@empty \href
  {http://dx.doi.org/#2} {doi:#2}\else \href {http://dx.doi.org/#2} {#1}\fi
  \endgroup}
\def\mn@eprint#1#2{\mn@eprint@#1:#2::\@nil}
\def\mn@eprint@arXiv#1{\href {http://arxiv.org/abs/#1} {{\tt arXiv:#1}}}
\def\mn@eprint@dblp#1{\href {http://dblp.uni-trier.de/rec/bibtex/#1.xml}
  {dblp:#1}}
\def\mn@eprint@#1:#2:#3:#4\@nil{\def\@tempa {#1}\def\@tempb {#2}\def\@tempc
  {#3}\ifx \@tempc \@empty \let \@tempc \@tempb \let \@tempb \@tempa \fi \ifx
  \@tempb \@empty \def\@tempb {arXiv}\fi \@ifundefined
  {mn@eprint@\@tempb}{\@tempb:\@tempc}{\expandafter \expandafter \csname
  mn@eprint@\@tempb\endcsname \expandafter{\@tempc}}}

\bibitem[\protect\citeauthoryear{Abadi et~al.,}{Abadi
  et~al.}{2015}]{tensorflow}
Abadi M.,  et~al., 2015, {TensorFlow}: Large-Scale Machine Learning on
  Heterogeneous Systems, \url {http://tensorflow.org/}

\bibitem[\protect\citeauthoryear{{Arons} \& {McCray}}{{Arons} \&
  {McCray}}{1970}]{Arons1969}
{Arons} J.,  {McCray} R.,  1970, \aplett, \href
  {https://ui.adsabs.harvard.edu/abs/1970ApL.....5..123A} {5, 123}

\bibitem[\protect\citeauthoryear{{Ba{\~n}ados} et~al.,}{{Ba{\~n}ados}
  et~al.}{2018}]{Banados2018}
{Ba{\~n}ados} E.,  et~al., 2018, \mn@doi [\nat] {10.1038/nature25180}, \href
  {https://ui.adsabs.harvard.edu/abs/2018Natur.553..473B} {553, 473}

\bibitem[\protect\citeauthoryear{{Barausse}, {Dvorkin}, {Tremmel}, {Volonteri}
  \& {Bonetti}}{{Barausse} et~al.}{2020}]{Barausse2020}
{Barausse} E.,  {Dvorkin} I.,  {Tremmel} M.,  {Volonteri} M.,   {Bonetti} M.,
  2020, arXiv e-prints, \href
  {https://ui.adsabs.harvard.edu/abs/2020arXiv200603065B} {p. arXiv:2006.03065}

\bibitem[\protect\citeauthoryear{{Bouwens}, {Illingworth}, {Oesch}, {Caruana},
  {Holwerda}, {Smit}  \& {Wilkins}}{{Bouwens} et~al.}{2015}]{Bouwens2015ems}
{Bouwens} R.~J.,  {Illingworth} G.~D.,  {Oesch} P.~A.,  {Caruana} J.,
  {Holwerda} B.,  {Smit} R.,   {Wilkins} S.,  2015, \mn@doi [\apj]
  {10.1088/0004-637X/811/2/140}, \href
  {https://ui.adsabs.harvard.edu/abs/2015ApJ...811..140B} {811, 140}

\bibitem[\protect\citeauthoryear{{Chardin}, {Haehnelt}, {Aubert}  \&
  {Puchwein}}{{Chardin} et~al.}{2015}]{Chardin2015}
{Chardin} J.,  {Haehnelt} M.~G.,  {Aubert} D.,   {Puchwein} E.,  2015, \mn@doi
  [\mnras] {10.1093/mnras/stv1786}, \href
  {https://ui.adsabs.harvard.edu/abs/2015MNRAS.453.2943C} {453, 2943}

\bibitem[\protect\citeauthoryear{Chollet et~al.}{Chollet et~al.}{2015}]{keras}
Chollet F.,  et~al., 2015, Keras, \url{https://keras.io}

\bibitem[\protect\citeauthoryear{{Ciardi}, {Ferrara}, {Marri}  \&
  {Raimondo}}{{Ciardi} et~al.}{2001}]{Ciardi2001}
{Ciardi} B.,  {Ferrara} A.,  {Marri} S.,   {Raimondo} G.,  2001, \mn@doi
  [\mnras] {10.1046/j.1365-8711.2001.04316.x}, \href
  {https://ui.adsabs.harvard.edu/abs/2001MNRAS.324..381C} {324, 381}

\bibitem[\protect\citeauthoryear{{Crain} et~al.,}{{Crain}
  et~al.}{2015}]{Crain2015}
{Crain} R.~A.,  et~al., 2015, \mn@doi [\mnras] {10.1093/mnras/stv725}, \href
  {https://ui.adsabs.harvard.edu/abs/2015MNRAS.450.1937C} {450, 1937}

\bibitem[\protect\citeauthoryear{{D'Aloisio}, {Upton Sanderbeck}, {McQuinn},
  {Trac}  \& {Shapiro}}{{D'Aloisio} et~al.}{2017}]{DAloisio2017}
{D'Aloisio} A.,  {Upton Sanderbeck} P.~R.,  {McQuinn} M.,  {Trac} H.,
  {Shapiro} P.~R.,  2017, \mn@doi [\mnras] {10.1093/mnras/stx711}, \href
  {https://ui.adsabs.harvard.edu/abs/2017MNRAS.468.4691D} {468, 4691}

\bibitem[\protect\citeauthoryear{{Dalal}, {White}, {Bond}  \&
  {Shirokov}}{{Dalal} et~al.}{2008}]{Dalal2008}
{Dalal} N.,  {White} M.,  {Bond} J.~R.,   {Shirokov} A.,  2008, \mn@doi [\apj]
  {10.1086/591512}, \href
  {https://ui.adsabs.harvard.edu/abs/2008ApJ...687...12D} {687, 12}

\bibitem[\protect\citeauthoryear{{DeGraf}, {Di Matteo}, {Khandai}  \&
  {Croft}}{{DeGraf} et~al.}{2012}]{Degraf2012}
{DeGraf} C.,  {Di Matteo} T.,  {Khandai} N.,   {Croft} R.,  2012, \mn@doi
  [\apjl] {10.1088/2041-8205/755/1/L8}, \href
  {https://ui.adsabs.harvard.edu/abs/2012ApJ...755L...8D} {755, L8}

\bibitem[\protect\citeauthoryear{{DeGraf}, {Di Matteo}, {Treu}, {Feng}, {Woo}
  \& {Park}}{{DeGraf} et~al.}{2015a}]{Volonteri2012}
{DeGraf} C.,  {Di Matteo} T.,  {Treu} T.,  {Feng} Y.,  {Woo} J.~H.,   {Park}
  D.,  2015a, \mn@doi [\mnras] {10.1093/mnras/stv2002}, \href
  {https://ui.adsabs.harvard.edu/abs/2015MNRAS.454..913D} {454, 913}

\bibitem[\protect\citeauthoryear{{DeGraf}, {Di Matteo}, {Treu}, {Feng}, {Woo}
  \& {Park}}{{DeGraf} et~al.}{2015b}]{DeGraf2015}
{DeGraf} C.,  {Di Matteo} T.,  {Treu} T.,  {Feng} Y.,  {Woo} J.~H.,   {Park}
  D.,  2015b, \mn@doi [\mnras] {10.1093/mnras/stv2002}, \href
  {https://ui.adsabs.harvard.edu/abs/2015MNRAS.454..913D} {454, 913}

\bibitem[\protect\citeauthoryear{{Di Matteo}, {Springel}  \& {Hernquist}}{{Di
  Matteo} et~al.}{2005}]{DiMatteo2005}
{Di Matteo} T.,  {Springel} V.,   {Hernquist} L.,  2005, \mn@doi [\nat]
  {10.1038/nature03335}, \href
  {https://ui.adsabs.harvard.edu/abs/2005Natur.433..604D} {433, 604}

\bibitem[\protect\citeauthoryear{{Di Matteo}, {Khandai}, {DeGraf}, {Feng},
  {Croft}, {Lopez}  \& {Springel}}{{Di Matteo} et~al.}{2012}]{DiMatteo2012}
{Di Matteo} T.,  {Khandai} N.,  {DeGraf} C.,  {Feng} Y.,  {Croft} R.~A.~C.,
  {Lopez} J.,   {Springel} V.,  2012, \mn@doi [\apjl]
  {10.1088/2041-8205/745/2/L29}, \href
  {https://ui.adsabs.harvard.edu/abs/2012ApJ...745L..29D} {745, L29}

\bibitem[\protect\citeauthoryear{{Di Matteo}, {Croft}, {Feng}, {Waters}  \&
  {Wilkins}}{{Di Matteo} et~al.}{2017}]{DiMatteo2017}
{Di Matteo} T.,  {Croft} R. A.~C.,  {Feng} Y.,  {Waters} D.,   {Wilkins} S.,
  2017, \mn@doi [\mnras] {10.1093/mnras/stx319}, \href
  {https://ui.adsabs.harvard.edu/abs/2017MNRAS.467.4243D} {467, 4243}

\bibitem[\protect\citeauthoryear{{Eide}, {Graziani}, {Ciardi}, {Feng},
  {Kakiichi}  \& {Di Matteo}}{{Eide} et~al.}{2018}]{Eide2018}
{Eide} M.~B.,  {Graziani} L.,  {Ciardi} B.,  {Feng} Y.,  {Kakiichi} K.,   {Di
  Matteo} T.,  2018, \mn@doi [\mnras] {10.1093/mnras/sty272}, \href
  {https://ui.adsabs.harvard.edu/abs/2018MNRAS.476.1174E} {476, 1174}

\bibitem[\protect\citeauthoryear{{Eide}, {Ciardi}, {Graziani}, {Busch}, {Feng}
  \& {Di Matteo}}{{Eide} et~al.}{2020}]{Eide2020}
{Eide} M.~B.,  {Ciardi} B.,  {Graziani} L.,  {Busch} P.,  {Feng} Y.,   {Di
  Matteo} T.,  2020, \mn@doi [\mnras] {10.1093/mnras/staa2774}, \href
  {https://ui.adsabs.harvard.edu/abs/2020MNRAS.498.6083E} {498, 6083}

\bibitem[\protect\citeauthoryear{{Fan} et~al.,}{{Fan} et~al.}{2019}]{Fan2019}
{Fan} X.,  et~al., 2019, \baas, \href
  {https://ui.adsabs.harvard.edu/abs/2019BAAS...51c.121F} {51, 121}

\bibitem[\protect\citeauthoryear{{Feng}, {Di Matteo}, {Croft}, {Tenneti},
  {Bird}, {Battaglia}  \& {Wilkins}}{{Feng} et~al.}{2015}]{Feng2015}
{Feng} Y.,  {Di Matteo} T.,  {Croft} R.,  {Tenneti} A.,  {Bird} S.,
  {Battaglia} N.,   {Wilkins} S.,  2015, \mn@doi [\apjl]
  {10.1088/2041-8205/808/1/L17}, \href
  {https://ui.adsabs.harvard.edu/abs/2015ApJ...808L..17F} {808, L17}

\bibitem[\protect\citeauthoryear{{Ferrarese} \& {Merritt}}{{Ferrarese} \&
  {Merritt}}{2000}]{Ferrarese2000}
{Ferrarese} L.,  {Merritt} D.,  2000, \mn@doi [\apjl] {10.1086/312838}, \href
  {https://ui.adsabs.harvard.edu/abs/2000ApJ...539L...9F} {539, L9}

\bibitem[\protect\citeauthoryear{{Field}}{{Field}}{1959}]{Field1959}
{Field} G.~B.,  1959, \mn@doi [\apj] {10.1086/146652}, \href
  {https://ui.adsabs.harvard.edu/abs/1959ApJ...129..525F} {129, 525}

\bibitem[\protect\citeauthoryear{{Finkelstein} et~al.,}{{Finkelstein}
  et~al.}{2019}]{Finkelstein2019}
{Finkelstein} S.~L.,  et~al., 2019, \mn@doi [\apj] {10.3847/1538-4357/ab1ea8},
  \href {https://ui.adsabs.harvard.edu/abs/2019ApJ...879...36F} {879, 36}

\bibitem[\protect\citeauthoryear{{Fletcher}, {Tang}, {Robertson}, {Nakajima},
  {Ellis}, {Stark}  \& {Inoue}}{{Fletcher} et~al.}{2019}]{Fletcher2019}
{Fletcher} T.~J.,  {Tang} M.,  {Robertson} B.~E.,  {Nakajima} K.,  {Ellis}
  R.~S.,  {Stark} D.~P.,   {Inoue} A.,  2019, \mn@doi [\apj]
  {10.3847/1538-4357/ab2045}, \href
  {https://ui.adsabs.harvard.edu/abs/2019ApJ...878...87F} {878, 87}

\bibitem[\protect\citeauthoryear{{Garaldi}, {Compostella}  \&
  {Porciani}}{{Garaldi} et~al.}{2019}]{Garaldi2019}
{Garaldi} E.,  {Compostella} M.,   {Porciani} C.,  2019, \mn@doi [\mnras]
  {10.1093/mnras/sty3414}, \href
  {https://ui.adsabs.harvard.edu/abs/2019MNRAS.483.5301G} {483, 5301}

\bibitem[\protect\citeauthoryear{{Giallongo} et~al.,}{{Giallongo}
  et~al.}{2015}]{Giallongo2015}
{Giallongo} E.,  et~al., 2015, \mn@doi [\aap] {10.1051/0004-6361/201425334},
  \href {https://ui.adsabs.harvard.edu/abs/2015A&A...578A..83G} {578, A83}

\bibitem[\protect\citeauthoryear{{Graziani}, {Maselli}  \& {Ciardi}}{{Graziani}
  et~al.}{2013}]{Graziani2013}
{Graziani} L.,  {Maselli} A.,   {Ciardi} B.,  2013, \mn@doi [\mnras]
  {10.1093/mnras/stt206}, \href
  {https://ui.adsabs.harvard.edu/abs/2013MNRAS.431..722G} {431, 722}

\bibitem[\protect\citeauthoryear{{Graziani}, {Ciardi}  \& {Glatzle}}{{Graziani}
  et~al.}{2018}]{Graziani2018}
{Graziani} L.,  {Ciardi} B.,   {Glatzle} M.,  2018, \mn@doi [\mnras]
  {10.1093/mnras/sty1367}, \href
  {https://ui.adsabs.harvard.edu/abs/2018MNRAS.479.4320G} {479, 4320}

\bibitem[\protect\citeauthoryear{{Hinton}, {Srivastava}, {Krizhevsky},
  {Sutskever}  \& {Salakhutdinov}}{{Hinton} et~al.}{2012}]{Hinton2012}
{Hinton} G.~E.,  {Srivastava} N.,  {Krizhevsky} A.,  {Sutskever} I.,
  {Salakhutdinov} R.~R.,  2012, arXiv e-prints, \href
  {https://ui.adsabs.harvard.edu/abs/2012arXiv1207.0580H} {p. arXiv:1207.0580}

\bibitem[\protect\citeauthoryear{{Huang}, {Di Matteo}, {Bhowmick}, {Feng}  \&
  {Ma}}{{Huang} et~al.}{2018}]{Huang2018}
{Huang} K.-W.,  {Di Matteo} T.,  {Bhowmick} A.~K.,  {Feng} Y.,   {Ma} C.-P.,
  2018, \mn@doi [\mnras] {10.1093/mnras/sty1329}, \href
  {https://ui.adsabs.harvard.edu/abs/2018MNRAS.478.5063H} {478, 5063}

\bibitem[\protect\citeauthoryear{{Hunter}}{{Hunter}}{2007}]{Hunter2007}
{Hunter} J.~D.,  2007, \mn@doi [Computing in Science and Engineering]
  {10.1109/MCSE.2007.55}, \href
  {https://ui.adsabs.harvard.edu/abs/2007CSE.....9...90H} {9, 90}

\bibitem[\protect\citeauthoryear{{Inayoshi}, {Visbal}  \& {Haiman}}{{Inayoshi}
  et~al.}{2019}]{Inayoshi2019}
{Inayoshi} K.,  {Visbal} E.,   {Haiman} Z.,  2019, arXiv e-prints, \href
  {https://ui.adsabs.harvard.edu/abs/2019arXiv191105791I} {p. arXiv:1911.05791}

\bibitem[\protect\citeauthoryear{{Khandai}, {Di Matteo}, {Croft}, {Wilkins},
  {Feng}, {Tucker}, {DeGraf}  \& {Liu}}{{Khandai} et~al.}{2015}]{Khandai2015}
{Khandai} N.,  {Di Matteo} T.,  {Croft} R.,  {Wilkins} S.,  {Feng} Y.,
  {Tucker} E.,  {DeGraf} C.,   {Liu} M.-S.,  2015, \mn@doi [\mnras]
  {10.1093/mnras/stv627}, \href
  {https://ui.adsabs.harvard.edu/abs/2015MNRAS.450.1349K} {450, 1349}

\bibitem[\protect\citeauthoryear{{Kormendy} \& {Ho}}{{Kormendy} \&
  {Ho}}{2013}]{Kormendy2013}
{Kormendy} J.,  {Ho} L.~C.,  2013, \mn@doi [\araa]
  {10.1146/annurev-astro-082708-101811}, \href
  {https://ui.adsabs.harvard.edu/abs/2013ARA&A..51..511K} {51, 511}

\bibitem[\protect\citeauthoryear{{Krawczyk}, {Richards}, {Mehta}, {Vogeley},
  {Gallagher}, {Leighly}, {Ross}  \& {Schneider}}{{Krawczyk}
  et~al.}{2013}]{Krawczyk2013}
{Krawczyk} C.~M.,  {Richards} G.~T.,  {Mehta} S.~S.,  {Vogeley} M.~S.,
  {Gallagher} S.~C.,  {Leighly} K.~M.,  {Ross} N.~P.,   {Schneider} D.~P.,
  2013, \mn@doi [\apjs] {10.1088/0067-0049/206/1/4}, \href
  {https://ui.adsabs.harvard.edu/abs/2013ApJS..206....4K} {206, 4}

\bibitem[\protect\citeauthoryear{{Kulkarni}, {Worseck}  \&
  {Hennawi}}{{Kulkarni} et~al.}{2019}]{Kulkarni2019}
{Kulkarni} G.,  {Worseck} G.,   {Hennawi} J.~F.,  2019, \mn@doi [\mnras]
  {10.1093/mnras/stz1493}, \href
  {https://ui.adsabs.harvard.edu/abs/2019MNRAS.488.1035K} {488, 1035}

\bibitem[\protect\citeauthoryear{{Ma}, {Ciardi}, {Eide}  \& {Helgason}}{{Ma}
  et~al.}{2018}]{Ma2018}
{Ma} Q.,  {Ciardi} B.,  {Eide} M.~B.,   {Helgason} K.,  2018, \mn@doi [\mnras]
  {10.1093/mnras/sty1806}, \href
  {https://ui.adsabs.harvard.edu/abs/2018MNRAS.480...26M} {480, 26}

\bibitem[\protect\citeauthoryear{{Madau}}{{Madau}}{2018}]{Madau2018}
{Madau} P.,  2018, \mn@doi [\mnras] {10.1093/mnrasl/sly125}, \href
  {https://ui.adsabs.harvard.edu/abs/2018MNRAS.480L..43M} {480, L43}

\bibitem[\protect\citeauthoryear{{Madau} \& {Dickinson}}{{Madau} \&
  {Dickinson}}{2014}]{Madau2014}
{Madau} P.,  {Dickinson} M.,  2014, \mn@doi [\araa]
  {10.1146/annurev-astro-081811-125615}, \href
  {https://ui.adsabs.harvard.edu/abs/2014ARA&A..52..415M} {52, 415}

\bibitem[\protect\citeauthoryear{{Madau} \& {Haardt}}{{Madau} \&
  {Haardt}}{2015}]{Madau2015}
{Madau} P.,  {Haardt} F.,  2015, \mn@doi [\apjl] {10.1088/2041-8205/813/1/L8},
  \href {https://ui.adsabs.harvard.edu/abs/2015ApJ...813L...8M} {813, L8}

\bibitem[\protect\citeauthoryear{{Madau}, {Meiksin}  \& {Rees}}{{Madau}
  et~al.}{1997}]{Madau1997}
{Madau} P.,  {Meiksin} A.,   {Rees} M.~J.,  1997, \mn@doi [\apj]
  {10.1086/303549}, \href
  {https://ui.adsabs.harvard.edu/abs/1997ApJ...475..429M} {475, 429}

\bibitem[\protect\citeauthoryear{{Madau}, {Haardt}  \& {Rees}}{{Madau}
  et~al.}{1999}]{Madau1999}
{Madau} P.,  {Haardt} F.,   {Rees} M.~J.,  1999, \mn@doi [\apj]
  {10.1086/306975}, \href
  {https://ui.adsabs.harvard.edu/abs/1999ApJ...514..648M} {514, 648}

\bibitem[\protect\citeauthoryear{{Maselli}, {Ciardi}  \& {Kanekar}}{{Maselli}
  et~al.}{2009}]{Maselli2009}
{Maselli} A.,  {Ciardi} B.,   {Kanekar} A.,  2009, \mn@doi [\mnras]
  {10.1111/j.1365-2966.2008.14197.x}, \href
  {https://ui.adsabs.harvard.edu/abs/2009MNRAS.393..171M} {393, 171}

\bibitem[\protect\citeauthoryear{{Mason}, {Naidu}, {Tacchella}  \&
  {Leja}}{{Mason} et~al.}{2019}]{Mason2019}
{Mason} C.~A.,  {Naidu} R.~P.,  {Tacchella} S.,   {Leja} J.,  2019, \mn@doi
  [\mnras] {10.1093/mnras/stz2291}, \href
  {https://ui.adsabs.harvard.edu/abs/2019MNRAS.489.2669M} {489, 2669}

\bibitem[\protect\citeauthoryear{{Matsuoka} et~al.,}{{Matsuoka}
  et~al.}{2018}]{Matsuoka2018}
{Matsuoka} Y.,  et~al., 2018, \mn@doi [\apj] {10.3847/1538-4357/aaee7a}, \href
  {https://ui.adsabs.harvard.edu/abs/2018ApJ...869..150M} {869, 150}

\bibitem[\protect\citeauthoryear{{Naidu}, {Forrest}, {Oesch}, {Tran}  \&
  {Holden}}{{Naidu} et~al.}{2018}]{Naidu2018}
{Naidu} R.~P.,  {Forrest} B.,  {Oesch} P.~A.,  {Tran} K.-V.~H.,   {Holden}
  B.~P.,  2018, \mn@doi [\mnras] {10.1093/mnras/sty961}, \href
  {https://ui.adsabs.harvard.edu/abs/2018MNRAS.478..791N} {478, 791}

\bibitem[\protect\citeauthoryear{{Nelson} et~al.,}{{Nelson}
  et~al.}{2018}]{Nelson2018}
{Nelson} D.,  et~al., 2018, \mn@doi [\mnras] {10.1093/mnras/stx3040}, \href
  {https://ui.adsabs.harvard.edu/abs/2018MNRAS.475..624N} {475, 624}

\bibitem[\protect\citeauthoryear{{Oesch}, {Bouwens}, {Illingworth}, {Labb{\'e}}
   \& {Stefanon}}{{Oesch} et~al.}{2018}]{Oesch2018}
{Oesch} P.~A.,  {Bouwens} R.~J.,  {Illingworth} G.~D.,  {Labb{\'e}} I.,
  {Stefanon} M.,  2018, \mn@doi [\apj] {10.3847/1538-4357/aab03f}, \href
  {https://ui.adsabs.harvard.edu/abs/2018ApJ...855..105O} {855, 105}

\bibitem[\protect\citeauthoryear{{Onoue} et~al.,}{{Onoue}
  et~al.}{2017}]{Onoue2017}
{Onoue} M.,  et~al., 2017, \mn@doi [\apjl] {10.3847/2041-8213/aa8cc6}, \href
  {https://ui.adsabs.harvard.edu/abs/2017ApJ...847L..15O} {847, L15}

\bibitem[\protect\citeauthoryear{{Parsa}, {Dunlop}  \& {McLure}}{{Parsa}
  et~al.}{2018}]{Parsa2018}
{Parsa} S.,  {Dunlop} J.~S.,   {McLure} R.~J.,  2018, \mn@doi [\mnras]
  {10.1093/mnras/stx2887}, \href
  {https://ui.adsabs.harvard.edu/abs/2018MNRAS.474.2904P} {474, 2904}

\bibitem[\protect\citeauthoryear{Pedregosa et~al.,}{Pedregosa
  et~al.}{2011}]{scikit-learn}
Pedregosa F.,  et~al., 2011, Journal of Machine Learning Research, 12, 2825

\bibitem[\protect\citeauthoryear{{Pfister}, {Volonteri}, {Dubois}, {Dotti}  \&
  {Colpi}}{{Pfister} et~al.}{2019}]{Pfister2019}
{Pfister} H.,  {Volonteri} M.,  {Dubois} Y.,  {Dotti} M.,   {Colpi} M.,  2019,
  \mn@doi [\mnras] {10.1093/mnras/stz822}, \href
  {https://ui.adsabs.harvard.edu/abs/2019MNRAS.486..101P} {486, 101}

\bibitem[\protect\citeauthoryear{{Planck Collaboration} et~al.,}{{Planck
  Collaboration} et~al.}{2020}]{PlanckCollaborationVI2018}
{Planck Collaboration} et~al., 2020, \mn@doi [\aap]
  {10.1051/0004-6361/201833910}, \href
  {https://ui.adsabs.harvard.edu/abs/2020A&A...641A...6P} {641, A6}

\bibitem[\protect\citeauthoryear{{Rees} \& {Setti}}{{Rees} \&
  {Setti}}{1970}]{Rees1969}
{Rees} M.~J.,  {Setti} G.,  1970, \aap, \href
  {https://ui.adsabs.harvard.edu/abs/1970A&A.....8..410R} {8, 410}

\bibitem[\protect\citeauthoryear{{Regan} \& {Haehnelt}}{{Regan} \&
  {Haehnelt}}{2009}]{Regan2009}
{Regan} J.~A.,  {Haehnelt} M.~G.,  2009, \mn@doi [\mnras]
  {10.1111/j.1365-2966.2008.14088.x}, \href
  {https://ui.adsabs.harvard.edu/abs/2009MNRAS.393..858R} {393, 858}

\bibitem[\protect\citeauthoryear{{Robertson}, {Ellis}, {Furlanetto}  \&
  {Dunlop}}{{Robertson} et~al.}{2015}]{Robertson2015}
{Robertson} B.~E.,  {Ellis} R.~S.,  {Furlanetto} S.~R.,   {Dunlop} J.~S.,
  2015, \mn@doi [\apjl] {10.1088/2041-8205/802/2/L19}, \href
  {https://ui.adsabs.harvard.edu/abs/2015ApJ...802L..19R} {802, L19}

\bibitem[\protect\citeauthoryear{{Shakura} \& {Sunyaev}}{{Shakura} \&
  {Sunyaev}}{1973}]{Shakura1973}
{Shakura} N.~I.,  {Sunyaev} R.~A.,  1973, \aap, \href
  {http://adsabs.harvard.edu/abs/1973A%26A....24..337S} {24, 337}

\bibitem[\protect\citeauthoryear{{Sijacki}, {Vogelsberger}, {Genel},
  {Springel}, {Torrey}, {Snyder}, {Nelson}  \& {Hernquist}}{{Sijacki}
  et~al.}{2015}]{Sijacki2015}
{Sijacki} D.,  {Vogelsberger} M.,  {Genel} S.,  {Springel} V.,  {Torrey} P.,
  {Snyder} G.~F.,  {Nelson} D.,   {Hernquist} L.,  2015, \mn@doi [\mnras]
  {10.1093/mnras/stv1340}, \href
  {https://ui.adsabs.harvard.edu/abs/2015MNRAS.452..575S} {452, 575}

\bibitem[\protect\citeauthoryear{{Tremmel}, {Governato}, {Volonteri}, {Pontzen}
   \& {Quinn}}{{Tremmel} et~al.}{2018}]{Tremmel2018}
{Tremmel} M.,  {Governato} F.,  {Volonteri} M.,  {Pontzen} A.,   {Quinn} T.~R.,
   2018, \mn@doi [\apjl] {10.3847/2041-8213/aabc0a}, \href
  {https://ui.adsabs.harvard.edu/abs/2018ApJ...857L..22T} {857, L22}

\bibitem[\protect\citeauthoryear{{VanderPlas}, {Connolly}, {Ivezic}  \&
  {Gray}}{{VanderPlas} et~al.}{2012}]{Vanderplas2012}
{VanderPlas} J.,  {Connolly} A.~J.,  {Ivezic} Z.,   {Gray} A.,  2012, in
  Proceedings of Conference on Intelligent Data Understanding (CIDU. pp 47--54
  (\mn@eprint {arXiv} {1411.5039}), \mn@doi{10.1109/CIDU.2012.6382200}

\bibitem[\protect\citeauthoryear{{Vanzella} et~al.,}{{Vanzella}
  et~al.}{2016}]{Vanzella2016}
{Vanzella} E.,  et~al., 2016, \mn@doi [\apj] {10.3847/0004-637X/825/1/41},
  \href {https://ui.adsabs.harvard.edu/abs/2016ApJ...825...41V} {825, 41}

\bibitem[\protect\citeauthoryear{{Virtanen} et~al.,}{{Virtanen}
  et~al.}{2020}]{virtanen2019}
{Virtanen} P.,  et~al., 2020, \mn@doi [Nature Methods]
  {10.1038/s41592-019-0686-2}, \href
  {https://ui.adsabs.harvard.edu/abs/2020NatMe..17..261V} {17, 261}

\bibitem[\protect\citeauthoryear{{Volonteri}, {Lodato}  \&
  {Natarajan}}{{Volonteri} et~al.}{2008}]{Volonteri2008}
{Volonteri} M.,  {Lodato} G.,   {Natarajan} P.,  2008, \mn@doi [\mnras]
  {10.1111/j.1365-2966.2007.12589.x}, \href
  {https://ui.adsabs.harvard.edu/abs/2008MNRAS.383.1079V} {383, 1079}

\bibitem[\protect\citeauthoryear{{Weinberger} et~al.,}{{Weinberger}
  et~al.}{2018}]{Weinberger2017}
{Weinberger} R.,  et~al., 2018, \mn@doi [\mnras] {10.1093/mnras/sty1733}, \href
  {https://ui.adsabs.harvard.edu/abs/2018MNRAS.479.4056W} {479, 4056}

\bibitem[\protect\citeauthoryear{{Worseck}, {Prochaska}, {Hennawi}  \&
  {McQuinn}}{{Worseck} et~al.}{2016}]{Worseck2016}
{Worseck} G.,  {Prochaska} J.~X.,  {Hennawi} J.~F.,   {McQuinn} M.,  2016,
  \mn@doi [\apj] {10.3847/0004-637X/825/2/144}, \href
  {https://ui.adsabs.harvard.edu/abs/2016ApJ...825..144W} {825, 144}

\bibitem[\protect\citeauthoryear{{Worseck}, {Davies}, {Hennawi}  \&
  {Prochaska}}{{Worseck} et~al.}{2019}]{Worseck2019}
{Worseck} G.,  {Davies} F.~B.,  {Hennawi} J.~F.,   {Prochaska} J.~X.,  2019,
  \mn@doi [\apj] {10.3847/1538-4357/ab0fa1}, \href
  {https://ui.adsabs.harvard.edu/abs/2019ApJ...875..111W} {875, 111}

\bibitem[\protect\citeauthoryear{{van der Walt}, {Colbert}  \&
  {Varoquaux}}{{van der Walt} et~al.}{2011}]{VanderWalt2011}
{van der Walt} S.,  {Colbert} S.~C.,   {Varoquaux} G.,  2011, \mn@doi
  [Computing in Science and Engineering] {10.1109/MCSE.2011.37}, \href
  {https://ui.adsabs.harvard.edu/abs/2011CSE....13b..22V} {13, 22}

\makeatother
\end{thebibliography}

\appendix

\bsp    
\label{lastpage}
\end{document}